\documentclass[runningheads]{llncs}
\usepackage{graphicx}
\pdfoutput=1
\usepackage{epsfig}
\usepackage{epigraph}
\usepackage{makecell}
\usepackage{tabularx} 
\usepackage{multirow}
\usepackage{xspace}
\usepackage{paralist}
\usepackage{libertine}
\usepackage{eurosym}
\usepackage{amssymb}
\usepackage{xcolor}
\usepackage{graphicx}
\usepackage{authblk}
\usepackage{cite}
\usepackage{pifont}
\usepackage{enumitem}
\definecolor[named]{MyRed}{cmyk}{0,0.84,0.8,0.19}
\definecolor[named]{MyGreen}{cmyk}{0.89,0,0.66,0.60}
\newcommand{\cmark}{\color{MyGreen} \ding{51}}%
\newcommand{\xmark}{\color{MyRed} \ding{55}}%

\usepackage{booktabs} 
\newcommand{\ra}[1]{\renewcommand{\arraystretch}{#1}}
\newcommand{\mycaption}[2]{\caption{\textbf{#1}. {#2}}}

\newcommand{\vheading}[1]{\vspace{0.05in}\noindent\textbf{#1}}

\newcommand{\eg}{\textit{e.g.,}\xspace}

 \newcommand{\article}{$\mathrsfso{G}$}
  \newcommand{\policy}{$\mathrsfso{P}$}
\newcommand{\policybold}{\textbf{$\mathrsfso{P}$}}

\usepackage[T1]{fontenc}
\DeclareFontFamily{T1}{calligra}{}
\DeclareFontShape{T1}{calligra}{m}{n}{<->s*[1.44]callig15}{}
\DeclareMathAlphabet\mathcalligra   {T1}{calligra} {m} {n}
\DeclareMathAlphabet\mathzapf       {T1}{pzc} {mb} {it}
\DeclareMathAlphabet\mathchorus     {T1}{qzc} {m} {n}
\DeclareMathAlphabet\mathrsfso      {U}{rsfso}{m}{n}

\usepackage{ifthen}
\newboolean{publicversion}
\setboolean{publicversion}{false}
\ifthenelse{\boolean{publicversion}}{
  \newcommand{\grumbler}[3]{}
}{
  \newcommand{\grumbler}[3]{\xspace\textcolor{#3}{\bf #1: #2}}
}

\begin{document}
\title{Analyzing GDPR Compliance Through the Lens of Privacy Policy}
\titlerunning{GDPR privacy policy}
%

\author{Jayashree Mohan \inst{1}\and
  Melissa Wasserman\inst{2}\and
Vijay Chidambaram \inst{1, 3} }
%
\authorrunning{J. Mohan et al.}

%

\institute{Department of Computer Science, University of Texas at
  Austin \and
  School of Law, University of Texas at Austin \and
VMWare Research, USA }

%
\maketitle              

\begin{abstract}
With the arrival of the European Union's General Data Protection
Regulation (GDPR), several companies are making significant changes to
their systems to achieve compliance. The changes range from modifying
privacy policies to redesigning systems which process personal
data. This work analyzes the privacy policies of large-scaled cloud
services which seek to be GDPR compliant. The privacy policy is the
main medium of information dissemination between the data controller
and the users. We show that many services that claim compliance today
do not have clear and concise privacy policies. We identify several
points in the privacy policies which potentially indicate
non-compliance; we term these GDPR vulnerabilities. We identify GDPR
vulnerabilities in ten cloud services. Based on our analysis, we
propose seven best practices for crafting GDPR privacy policies. 
\end{abstract}

\section{Introduction}

Security, privacy, and protection of personal data have become complex 
and absolutely critical in the Internet era. Large scale cloud infrastructures 
like Facebook have focused on scalability as one of the primary goals 
(as of 2019, there are 2.37 billion monthly active users on facebook~\cite{fb-users}),
leaving security and privacy on the backseat. This is evident from the gravity
of personal data breaches reported over the last decade. For instance, the 
number of significant data breaches at U.S. businesses, government agencies, and other organizations was over 1,300 in 2018, as compared to 
fewer than 500, ten years ago~\cite{breaches-count}. The magnitude of impact of
such breaches are huge; for example, the Equifax breach~\cite{equifax} compromised 
the financial information of $\sim$145 million consumers. In response to 
the alarming rise in the number of data breaches, the
European Union (EU) adopted a comprehensive privacy regulation called
the General Data Protection Regulation (GDPR)~\cite{gdpr-regulation}.

At the core of GDPR is a new set of rules and regulations, aimed at providing the 
citizens of the EU, more control over their personal data.
Any company or organization operational in the EU and dealing with the personal
data of EU citizens is legally bound by the laws laid by GDPR.
GDPR-compliant services must ensure that personal data is collected legally 
for a specific purpose, and are obliged to protect it from misuse and exploitation; 
failure to do so, might result in hefty penalties for the company. As of Jan 2019, 
91 reported fines have been imposed under the new GDPR regime
~\cite{gdpr-fines}. The magnitude of fine imposed varies by the severity of
non-compliance. For instance, in Germany, a \euro 20,000 fine was imposed on a company 
for failing to hash employee passwords, resulting in a security breach. Whereas 
the French data protection authority fined Google  \euro 50 million for not properly 
disclosing to users how data is collected across its services to present personalized advertisements. A series of lawsuits and fines have now forced companies to 
take a more  \textit{privacy-focused} future for their services~\cite{fb-future}.

While our prior work examined how GDPR affects the design and
operation of Internet companies~\cite{gdpr-sins} and its impact on
storage systems~\cite{gdpr-storage}, this work focuses on a third
dimension : privacy policies (PP).  A privacy policy is a statement or
a legal document (in privacy law) that discloses the ways a party
gathers, uses, discloses, and manages a customer or client's
data~\cite{flavian2006consumer}. The key
to achieving transparency, one of the six fundamental data protection
principles laid out by GDPR, is a clear and concise PP that informs
the users how their data is collected, processed, and controlled. We
analyze the privacy policies of ten large-scale cloud services that
operate in the EU and claim to be GDPR-compliant; we identify several
\emph{GDPR vulnerabilities}, points in the PP that could potentially
lead to non-compliance with GDPR. Some of the vulnerabilities we
identify are clear-cut non-compliance (\eg not providing details about
the Data Protection Officer), while others lie in grey areas and are
up for interpretation. However, based on the prior history of fines
levied on charges of GDPR non-compliance~\cite{gdpr-fines}, we believe there is
a strong chance that all identified vulnerabilities may lead to
charges.

Our analysis reveals that most PP are not clear and concise, sometimes
exploiting the vague technical specifications of GDPR to their
benefit. For instance, Bloomberg, a software tech company states in
its PP that \textit{"Bloomberg may also disclose your personal
  information to unaffiliated third parties if we believe in good
  faith that such disclosure is necessary [...]"}, with no mention of
who the third-parties are and how to object to such disclosure and
processing. Furthermore, we identify several vulnerabilities in the PP
that indicate potential non-compliance with GDPR. First, many services
exhibit all-or-none behaviors with respect to user controls over data
oftentimes requiring withdrawal from the service to enable deletion of
any information, or to raise objections to data processing. Second,
most controllers bundle the purposes for data collection and
processing amongst various entities.  They collect multiple categories
of user data, each on a different platform and state a bunch of
purposes for which they, or their Affiliates could use this data. We
believe this is in contradiction to GDPRs goals of attaching a purpose
to every piece of collected personal information.

Based on our study, we propose seven policy recommendations that a
GDPR-compliant company should address in their PP.  The proposed
policy considerations correspond to the categories of data collected,
their purpose, the lawfulness of processing them, etc. We accompany
each consideration with the GDPR article that necessitates it and
where applicable, provide an example of violation of this policy by
one of the systems under our study.

Our analysis is not without limitations. First, while we studied a
wide category of cloud-services ranging from social media to
education, our study is not exhaustive; we do not analyze categories
like healthcare, entertainment, or government services. Second, we do
not claim to identify all vulnerabilities in each PP we
analyzed. Despite these limitations, our study contributes useful
analyses of privacy policies and guidelines for crafting
GDPR-compliant privacy policies.

\section{GDPR and Privacy Policy }
\label{sec-bgk}

\vheading{GDPR}
The General Data Protection Regulation (GDPR) came into effect on May 
25th 2018 as the legal framework that sets guidelines for the collection 
and processing of personal information of people in the European Union 
(EU)~\cite{gdpr-regulation}. The primary goal of GDPR is to ensure 
protection of personal data by vesting the control over data in the users 
themselves. Therefore, the \emph{data subject} (the person whose 
personal data is collected) has the power to demand companies to reveal 
what information they hold about the user, object to processing his data, 
or request to delete his data held by the company. GDPR puts forth 
several laws that a data collector and processor must abide by; such 
entities are classified either as \emph{data controller}, the entity that 
collects and uses personal data, 
 or as a \emph{data processor}, the entity that processes personal data on 
behalf of a data 
controller, and the regulations may vary for the two entities. 

\vheading{Key policies of GDPR}. The central focus of GDPR is to provide 
the data subjects extensive control over their personal data 
collected by the controllers. Companies that wish to stay GDPR-compliant 
must take careful measures to ensure protection of user data by 
implementing state-of-the-art techniques like 
pseudonymization and encryption. They should also provide the data 
subjects with ways to retrieve, delete, and raise objections to the use of 
any information pertaining to them. Additionally, the companies should 
appoint supervisory authorities like the Data Protection Officer (DPO) to 
oversee the company's data protection strategies and must notify data 
breaches within 72 hours of first becoming aware of it. Failure to comply to 
GDPR can result in hefty fines; up to 4\% of the annual global turnover of 
the company.

\vheading{Impact of GDPR}. 
Several services shut down completely, while others blocked access to the 
users in the European Union(EU) in response to GDPR. For instance, the 
need for infrastructural changes led to the downfall of several multiplayer 
games in the EU, including Uber Entertainment's Super Monday Night 
Combat and Gravity Interactive's Ragnarok Online~\cite{gaming-shutdown}, 
whereas the changes around user consent for data processing resulted in 
the shut down of advertising companies like Drawbridge~\cite{drawbridge-shut}. Furthermore, 91 reported fines have been imposed under the new 
GDPR regime as of January 2019, with charges as high as \euro 50 
million~\cite{gdpr-fines}.

\vheading{GDPR and privacy policy}.
A privacy policy is a statement or a legal document (in privacy law) that 
discloses some or all of the 
ways a party gathers, uses, discloses, and manages a customer or client's 
data~\cite{wiki-pp, flavian2006consumer}. 
It is the primary grounds for transparent data processing requirements set 
forth by GDPR. GDPR \article 12  sets the ground for transparency, one of 
the six fundamental principles of GDPR. It states that any information or 
communication to the users must be \textit{concise, transparent, 
intelligible and in an easily accessible form, using clear and plain 
language}. The main objective of this article is to ensure that users are 
aware of how their data is collected, used, and processed. Therefore, the 
first step towards GDPR compliance at the controllers is updating the 
privacy policy, which is the primary information notice board between the 
controller and the customer.

\section {Best Practices for GDPR compliant Privacy Policies}
\label{sec-bestprac}


GDPR has six general data protection principles (transparency; purpose  limitation;  
data  minimization;  accuracy; storage limitation; and confidentiality)  with  data  
protection  by  design  and  default  at  the  core. The first step to implementing 
these data-protection principles is to conceptualize an accurate privacy policy at the data controller.

Privacy policy documents issued by data controllers are oftentimes overlooked by customers either 
because they are too lengthy and boring, or contain too many technical jargons.  For instance, Microsoft's privacy policy is 
56 pages of text~\cite{ms-policy}, Google's privacy policy spans 27 pages of textual content~\cite{google-policy}, and
Facebook's data policy document is 7 pages long~\cite{fb-policy}. A Deloitte survey of 2,000 consumers in the U.S found that 91\% 
of people consent to legal terms and service conditions without reading them~\cite{policy-survey}. 

Privacy policies of GDPR-compliant systems must be specific about the sharing and distribution of user data to third-
parties, with fine-grained access control rights to users. On the contrary, Apple iCloud's  privacy 
policy reads as follows~\cite{icloud-policy} : \textit{[...] You acknowledge and agree that Apple may, 
without liability to you, access, use, preserve and/or disclose your Account information and Content 
to law enforcement authorities, government officials, and/or a third party, as Apple believes is 
reasonably necessary or appropriate [...] }. While this contradicts the goals of GDPR, this 
information is mentioned on the 11th page of a 20 page long policy document, which most 
customers would tend to skip.

These observations put together, emphasizes the need for a standardized privacy-policy document 
for GDPR-compliant systems. We translate GDPR articles into precise questions that a user must 
find answer to, while reading any privacy policy. An ideal privacy policy for a GDPR-complaint system should at the least, answer all of the following questions prefixed with \policy. The 
GDPR law that corresponds to the question is prefixed with \article.

 \vheading{(\policybold 1) : Processing Entities}. Who collects the user data and who are the users of data (\article 5(1)B, 6, 21).  
 
 The PP of a GDPR-compliant controller must precisely state the sources of data, and with whom the collected data is shared. While many controllers vaguely state that they \textit{"may share the data with third-parties"}, GDPR requires specifying who the third parties are, and for what purpose they would use this data.
 

 \vheading{(\policybold 2) : Data Categories}.  What personally identifiable data is collected (\article 14, 20)
 
 The controller must clearly state the attributes of personal data (name, email, phone number, IP etc)  being collected or at the least, categories of these attributes. All the PP we studied fairly addresses this requirement.

 \vheading{(\policybold 3) : Retention}.  When will the collected data expire and be deleted (\article 5(1)E, 13, 17)
 
GDPR requires that the controller attach a retention period or a basis for determining this retention period to every category of personal data collected for a specific purpose. Such retention periods or policies must be mentioned straight up in the PP.  Apple's PP for instance has no mention of how long the collected data will reside in their servers~\cite{apple-policy}. It also has no details on whether the user data will ever be deleted after its purpose of collection is served.
 
 \vheading{(\policybold 4) : Purpose }. Why is the data being collected \article 5(1)B)
 
 Purpose of data collection is one of the main principles of data protection in GDPR. The PP must therefore clearly state the basis for collection of each category of personal data and the legal basis for processing it. The controller should also indicate if any data is obtained from third-parties and the legal basis for processing such data.

 \vheading{(\policybold 5) : User Controls}. How can the user request the following
	\begin{enumerate}[label=\alph*), leftmargin=1cm]
	\item All the personal data associated with the user along with its source, purpose, TTL, the list 
of third-parties to which it has been shared etc (\article 15)
	\item Raise objection to the use of any attribute of their personal data (\article 21)
	\item Personal data to be deleted without any undue delay (\article 17)
	\item Personal data to be transferred to a different controller (\article 20)
	\end{enumerate}

Not all PP explicitly state the user's rights to access and control their personal data. For instance,  Uber has no option to request for deletion of user travel history, without having to deactivate the account. 

 \vheading{(\policybold 6) :  Data Protection}. Does the controller take measures to ensure safety and protection of data
	\begin{enumerate}[label=\alph*), leftmargin=1cm]
	\item By implementing state-of-the-art techniques such as encryption or  pseudonymization (\article 25, 32)
	\item By logging all activities pertaining to user data (\article 30)
	\item By ensuring safety measures when processing outside the EU (\article 3)
	\end{enumerate}
	
	GDPR puts the onus of data protection by design and default on the data controller. Additionally, irrespective of the location of processing data at the controller, they should be clear about the data protection guarantees provided when processed outside he EU. Additionally, the PP must contain the contact details of the data protection officer (DPO) or appropriate channels to request, modify or delete their information. 
	
 \vheading{(\policybold 7) : Policy Updates}. Does the controller notify users appropriately when changes are made to the privacy policy (\article 14)
 
 The transparency principle of GDPR advocates that the users must be notified and be given the chance to review and accept the new terms, whenever changes are made to policies. On the contrary, many services simply update the date of modification in the policy document rather than taking measures to reasonably notify the users (for eg., using email notifications).

\section{Case Studies}
\label{sec-case}

\begin{table*}[!t]
  \small
  \centering
  \ra{1.3}
  \begin{tabular}{l| c | c | c | c | c | c | c |}
    \toprule[1.2pt]
     Cloud Service  & \policy 1 &  \policy 2 & \policy 3& \policy 4& \policy 5  &  \policy 6 & \policy 7\\
     & Processing & Data &  Retention & Purpose & Controls & Protection & Updates \\
    \midrule
    Bloomberg & \xmark & \cmark & \xmark & \cmark& \xmark& \xmark & \xmark  \\
    Onavo & \xmark & \cmark & \xmark & \cmark& \xmark& \xmark & \cmark  \\
    Instagram & \xmark & \cmark & \xmark & \cmark& \cmark& \xmark & \cmark  \\
    Uber & \xmark & \cmark & \cmark & \cmark& \xmark& \xmark & \cmark  \\
     edx & \cmark & \cmark & \cmark & \cmark& \xmark& \xmark & \xmark  \\
     \midrule
     Snapchat & \cmark & \cmark & \cmark & \cmark& \cmark& \cmark & \xmark  \\
    iCloud & \xmark & \cmark & \cmark & \cmark& \cmark& \cmark & \cmark  \\
     \midrule
     Whatsapp & \cmark & \cmark & \cmark & \cmark& \cmark& \cmark & \cmark\\
     FlyBe Airlines & \cmark & \cmark & \cmark & \cmark& \cmark& \cmark & \cmark  \\
     Metro bank & \cmark & \cmark & \cmark & \cmark& \cmark& \cmark & \cmark  \\
    \bottomrule[1.2pt]
  \end{tabular}
\vspace{1.5pt}
  \mycaption{GDPR vulnerabilities}{The table shows GDPR vulnerabilities across 10 cloud services. }
  \label{tbl-vulnerabilities}
\vspace{-0.3in}  
\end{table*}

In this section we present the case study of ten large-scale cloud services that are operational in 
the EU. We analyze various categories of applications and services including three social 
media applications like Whatsapp and Instagram to financial institutions like Metro bank. We study the privacy policies of each of these services and 
identify GDPR vulnerabilities that could lead to potential GDPR non-compliance.
Table~\ref{tbl-vulnerabilities} categorizes companies in the descending order of 
GDPR vulnerabilities. The discussion below is grouped by the type of commonly 
observed vulnerabilities. 

\vheading{Unclear data sharing and processing policies}. 
Instagram, a photo and video-sharing social networking service owned by 
Facebook Inc discloses user information to all its Affiliates ( the Facebook 
group of companies),  who can use the information with no specific user consent~\cite{instagram-policy}. The way 
in which Affiliates use this data is claimed to be \textit{"under reasonable confidentiality terms"}, which is vague. For instance, it is unclear whether a mobile number that is marked private 
in the Instagram account, is shared with, and used by Affiliates. This can count towards 
violation of purpose as the mobile number was collected primarily for account creation 
and cannot be used for other purposes without explicit consent. Additionally, Instagram 
says nothing about the user's right to object to data processing by Affiliates  or third-parties. 
It's PP says \textit{"Our Service Providers will be given access to your information as is 
\textbf{reasonably necessary} to provide the Service under reasonable confidentiality 
terms"}. Uber on the other hand, may provide collected information to its vendors, 
consultants, marketing partners, research firms, and other service providers or business 
partners, but does not specify how the third parties would use this information~\cite{uber-policy}. On similar 
grounds, iCloud's PP vaguely states that information may be shared with third-parties, but 
does not specify who the third-parties are, and how to opt-out or object to such sharing~\cite{icloud-policy}.. 
Similarly Bloomberg is vague about third-party sharing and says, \textit{"Bloomberg may 
also disclose your personal information to unaffiliated third parties if we believe in good 
faith that such disclosure is necessary [...]"}~\cite{bloomberg-policy}.

\vheading{Vague data retention policies}. Instagram does not guarantee that user data is 
completely deleted from its servers when a user requests for deletion of personal 
information. Data can remain viewable in cached and archived pages of the service. 
Furthermore Instagram claims to store the user data for a  \textit{"reasonable"} amount of 
time for \textit{"backup"}, after account deletion, with no justification of why it is necessary, 
and whether they will continue to use the backup data for processing. Other companies 
including Bloomberg and Onavo do not specify a retention period, vaguely specifying that 
\textit{personal information is retained for as long as is necessary for the purpose for which 
it is collected}~\cite{onavo-policy, bloomberg-policy}.

\vheading{Un-reasonable ways of notifying updates to privacy policy}. Changes to PP 
should be notified to all users in a timely manner and users must be given the chance to 
review and accept the updated terms. However, edX, Bloomberg, and Snapchat would 
simply \textit{"label the Privacy Policy as "Revised (date)[...]. By accessing the Site after any 
changes have been made, you accept the modified Privacy Policy and any changes 
contained therein"}~\cite{edx-policy, bloomberg-policy, snap-policy}. This is un-reasonable as it is easy to miss such notifications, and a 
better way of notifying users is by sending an email to review the updated policy.

\vheading{Weak data protection policies}. GDPR  \article 37 requires the controller to 
publish contact details of the data protection officer (DPO). The privacy policies of 
Instagram, Facebook, Bloomberg and edX have no reference to who the DPO is, or how to 
contact them. Similarly, while most cloud services assure users that their data processing will abide 
by the terms in the PP irrespective of the location of processing, services like Onavo take a 
laidback approach. It simply states that they \textit{"may process your information, including 
personally identifying information, in a jurisdiction with different data protection laws than 
your jurisdiction"}, with nothing said about the privacy guarantees in cases of such 
processing. Some other services like Uber, state nothing about data protection techniques 
employed or international transfer policies.

\vheading{No fine-grained control over user data}. The edX infrastructure does not track 
and index user data at every place where the user volunteers information of the site. 
Therefore, they claim that, \textit{"neither edX nor Members will be able to help you locate 
or manage all such instances."}. Similarly, deleting user information does not apply to 
\textit{"historical activity logs or archives unless and until these logs and data naturally ?
age-off? the edX system"}. It is unclear if such data continues to be processed after a user 
has requested to delete his information. Similarly, Uber requires the user to deactivate their 
account to delete personal information from the system. Moreover, if a user objects to the 
usage of certain personal information, , \textit{" Uber may 
continue to process your information notwithstanding the objection to the extent permitted 
under GDPR"}. It is unclear to what extent and on what grounds Uber can ignore the 
objections raised by users. While most services provide a clear overview the rights user 
can exercise and the ways of doing so by logging into their service, Onavo simply states, 
\textit{"For assistance with exercising rights, you can contact us at support@onavo.com"}. 
It does not specify what kind of objections can be raised, what part of the personal 
information can be deleted, etc.

\subsection{A good privacy policy}
Flybe is a British airlines whose privacy policy was by far the most precise and clear document of all the 
services we analyzed~\cite{flybe-policy}, probably because it's based in the EU. 
Nonetheless, the effort put by Flybe into providing all necessary information pertaining to 
the collection and use of customer's personal data is an indicator of its commitment to 
GDPR-compliance. For instance, Flybe clearly categorizes types of user information 
collected, along with a purpose attached to each category. While most of the services we 
analyzed claim to simply share information with third-parties as necessary, Flybe 
enumerates each of its associated third-parties, the specifics of personal data shared with 
them, the purpose for sharing and a link to the third-parties privacy policy. In cases where it 
is necessary to process user data outside of EU, Flybe ensures a similar degree of 
protection as in the EU. We believe that a PP as clear as the one employed by Flybe, 
enables users to gain a fair understanding of their data and their rights over collected data. 
The level of transparency and accountability demonstrated by this PP is an indicator of right 
practice for GDPR-compliance. 

\subsection{Summary}
The major GDPR vulnerabilities we identify in large-scale cloud services can be summarized as follows. 

\vheading{All or nothing}. Most companies have rolled out new policies and products to 
comply with GDPR, but  those policies don't go far enough. In particular, the way 
companies obtain consent for the privacy policies is by asking users to check a box in order to 
access services. It is a widespread practice for online services, but  it forces users into an 
all-or-nothing choice, a violation of the GDPR's provision around particularized consent 
and fine-grained control over data usage. There's a lawsuit against Google and Facebook 
for a similar charge~\cite{gdpr-lawsuit}.

This behavior extends to other types of user rights that GDPR advocates. For instance, 
GDPR vests in the users the right to object to the use of a part or all of their personal data, 
or delete it. Most controllers however, take the easy approach and enable these knobs only 
if they user un-registers for their service. This approach is not in the right spirit of GDPR.

\vheading{Handwavy about data protection}. GDPR requires controllers to adopt internal 
policies and implement measures which meet in particular ,the principles of data protection 
by design and default. However, many cloud services seem to dodge the 
purpose by stating that in spite of the security measures taken by them (they do not specify 
what particular measures are taken), the user data may be accessed, disclosed, altered, or 
destroyed. Whether this is non-compliance is a debatable topic, however, the intent of 
GDPR \article 24 and 25  is to encourage controllers to implement state of the art data 
protection techniques.

\vheading{Purpose Bundling}. Most controllers bundle the purposes for data collection and 
processing amongst various entities. They collect multiple categories of user data, each on 
a different platform and state a bunch of purposes for which they, or their Affiliates could 
use this data. Although this might not be explicit non-compliance, it kills GDPR's notion of a 
purpose attached to every unit of user data collected. 

\vheading{Unreasonable Privacy Policy Change Notifications}. Privacy policy being the 
binding document based on which a user consents to using a service, any changes to the 
policy must be notified to the user in a timely and appropriate manner. This may include 
sending an email to all registered users, or in case of a website, placing a notification pop-
up without reading and accepting which, the user cannot browse further. However, many 
services we analyzed have unreasonable update policies, where in they simply update the 
last modified date in the privacy policy and expect the user to check back frequently.

\subsection{User experiences with exercising GDPR rights in the real world} 
Privacy policies provide an overview of techniques and strategies employed by the 
company to be GDPR-compliant, including the rights that the users could exercise over 
their data. While no lawsuit can be filed against a company unless there is a proof for 
violation of any of the GDPR laws claimed in the PP, this section is an account of some 
user's attempts to exercise the rights claimed in the PP.

A user of Pokemon Go raised an objection to processing her personal data, and  to stop 
using her personal data for marketing and promotional purposes, both of which are listed 
under the user's rights and choices in Pokemon Go's PP. The response from the controller 
however, was instructions on how to delete the user account~\cite{pokemon-twitter}. In another incident, Carl 
Miller, Research Director at the Centre for the Analysis of Social Media requested an 
unnamed company to return all personal data they hold about him (which is a basic right 
GDPR provides to a data subject). However, the company simply responded that they are 
not the controller for the data he was asking for~\cite{carl-twitter}. Adding on to this, when a 
user requests for personal information, the company requires him to specify what data he 
needs~\cite{carl-twitter2}. This is not in the right spirit of GDPR because, a user does not 
know what data a controller might have. This violates the intent of GDPR because the main 
idea is to give users a better idea of what data is held about them. 

These real experiences of common people show that GDPR has a long way to go, to 
achieve its goal of providing users with knowledge and control over all their personal 
information collected and processed by various entities.

\section{Discussion}
\label{sec-discussion}

The negative responses received by users trying to exercise their GDPR rights, and the shut 
down of several services in the European Union(EU) in response to GDPR, motivated us to 
analyze the root cause of this behavior. We identify two prominent reasons.

First, some companies do not have well informed policies for sharing the collected data across 
third-parties, or they rely completely on information from third-parties for their data. Second, their 
infrastructure does not support identifying, locating, and packaging user data in response to 
user queries. While the former can be resolved by ensuring careful data sharing policies, the 
latter requires significant reworking of backend infrastructure. Primarily, the need for 
infrastructural changes led to the downfall of several multiplayer games in the EU, including 
Uber Entertainment's Super Monday Night Combat, Gravity Interactive's Ragnarok Online 
and Dragon Saga and Valve's entire gaming community~\cite{gaming-shutdown}. In this context, we identify 4 primary 
infrastructural changes that a backend storage system must support in order to be GDPR-complaint~\cite{gdpr-storage} and suggest possible solutions in each case. 

\vheading{Timely deletion}. Under GDPR, no personal data can be retained 
for an indefinite period of time. Therefore, the storage system should 
support mechanisms to associate time-to-live (TTL) counters for 
personal data, and then automatically erase them from all internal 
subsystems in a timely manner. One way to efficiently allow deletion is to maintain a 
secondary index on TTL like time series databases~\cite{faloutsos1994fast}.

\vheading{Indexing via Metadata}. Several articles of GDPR require efficient access to 
groups of data based on certain 
attributes.  While traditional databases natively offer this ability via secondary indices, not all 
storage systems have
efficient or configurable support for this capability. Several research in the past have 
explored building efficient multi-index stores. The common technique used in multi-index 
stores is to utilize redundancy to partition each full copy of the data by a different 
key~\cite{tai2016replex, sivathanu2019instalytics}. 

\vheading{Monitoring and Logging}. GDPR allows the data subject to query the usage 
pattern of their data. Therefore, the storage system needs an audit trail of both its internal 
actions and external interactions. One way to tackle this problem is to use fast non-volatile 
memory devices like 3D Xpoint to store logs. Efficient auditing may also be achieved through 
the use of eidetic systems. For example, Arnold is able to remember 
past state with only 8\% overhead~\cite{eidetic-systems}.

\vheading{Access Control and Encryption}. As GDPR aims to limit access to personal 
data to only permitted entities, for established purposes, and for 
predefined duration of time, the storage system must support
fine-grained and dynamic access control. An effective way of doing this is to break down user 
data, and encrypt each attribute using a different public key; an approach commonly termed as
Key-Policy Attribute-Based Encryption (KP-ABE)~\cite{goyal2006attribute}. 

\section{Conclusion}
\label{sec-conc}

We analyze the privacy policies of ten large-scale cloud services, identifying vulnerabilities that could potentially result in GDPR non-compliance. 
While our study shows that many PP are far from clear, we also provide real world examples to show that exercising user rights claimed in PP is not an easy task. Additionally, we propose seven recommendations that a PP should address, to be close to GDPR-compliance.

With the growing relevance of privacy regulations around the 
world, we expect this paper to trigger interesting conversations around the need for clear and concrete GDPR-compliant privacy policies. We are keen to extend our effort to 
engage the storage community in addressing the 
research challenges in alleviating the identified GDPR vulnerabilities, by building better infrastructural support where necessary.


{
\bibliographystyle{splncs04}
\bibliography{all}
}

\end{document}